# A Novel Recurrent Adaptive Backstepping Optimal Control Strategy for a Single Inverted Pendulum System

Mohammad Sarbaz

Electrical and Electronic Engineering Department, Shahed University, Tehran, Iran
Corresponding author. Tel.: +98 21 51212029.
E-mail addresses: mohammad.sarbaz@shahed.ac.ir (M. Sarbaz).

**Abstract-**In this paper, a novel recurrent adaptive backstepping optimal control strategy for a single inverted pendulum system is studied. By this method, an inverted pendulum is stabilized using projection recurrent neural network-based adaptive backstepping control (PRNN-ABC). The inverted pendulum is a popular nonlinear system that is used in both industry and academic and is applied various control approaches since it has many applications. Here, first of all, the backstepping control laws are investigated based on the nonlinear dynamic model of the system. Second, by considering control constrains and performance index, the constrained optimization problem is formulated. Later, the optimization problem will be converted to a constrained quadratic problem (QP). To study the recurrent neural network (RNN) according to the Karush-Kuhn-Tucker (KKT) optimization conditions and the variational inequality, the dynamic model of the RNN will be derived. At last, the stability analysis of the system is studied using Lyapunov function.

*Keywords*- Single Inverted Pendulum, backstepping optimal control, recurrent adaptive neural network, variational inequality.

## Introduction

A single inverted pendulum is a pendulum that has its center of mass above its pivot point. It is unstable and without additional help and any external force of disturbance will be fallen over and crashed down easily. It can be suspended stably in this inverted position by using a control system to monitor the angle of the pole and move the pivot point horizontally back under the center of mass when it starts to fall over, keeping it balanced. The inverted pendulum is a classic and wide-spread problem in dynamics and control theory and is used as a benchmark for testing control strategies. This system has a potential capability to be tasted and applied various approaches due to its accessibility and wide-spread. usually, high-speed and perfect performance, and a reduction in friction and, as a result, an increase in lifespan are expected. This concept has been obtained in both laboratory and industry.

Many researches have been done regarding the inverted pendulum and many control algorithms have been applied to this system to stabilize and obtain smooth output results [1], [2], and [3]. For instance, identification of the inverted pendulum to establish the dynamic model of inverted pendulum [4], using LQR algorithm for the balancing control of both pendulums and position control [5], applying adaptive feedback linearization control and identifying the system using rough neural network [6], or considering a novel two degree of freedom (2-DOF) fractional control strategy based on 2-loop topology for inverted pendulum [7].

Inverted pendulum implemented with the pivot point stands on a surface that move horizontally under control of an electronic servo system., named a cart and pole apparatus. Most applications limit the pendulum to 1 degree of freedom by affixing the pole to an axis of rotation. Whereas a normal pendulum is stable when hanging downwards, an inverted pendulum is inherently unstable, and must be actively balanced to remain upright; this can be done by applying a torque at the pivot point, or moving the pivot point horizontally as part of a feedback system, changing the rate of rotation of a mass mounted on the pendulum on an axis parallel to the pivot axis and thereby generating a net torque on the pendulum, or by oscillating the pivot point vertically. A simple demonstration of moving the pivot point in a feedback system is achieved by balancing an upturned broomstick on the end of one's finger. Since having the exact dynamic model of the inverted pendulum is a hard task and by these days' novel approaches it is not efficient to work with dynamic model directly, different identification methods like neural networks are used widely [8] and [9]. Additionally, some classical control strategy like PI-PD controllers [10] or linear parameter–varying controller [11] are applied to the identified model of the inverted pendulum.

One of the most common method in control systems is backstepping control that by using an analytical and step-by-step approach is able to realize various operation modes. In this control approach, regarding different dynamic subsections, the control laws are computed by defining virtual variables and stabilizing them by the Lyapunov stability theory. Some researches and methods are proposed in this area sine years ago [12] and [13]. In [14], for an autonomous lane-keeping system, a backstepping control method with an augmented observer is studied. In [15], an adaptive backstepping control is applied to a parametric strict-feedback nonlinear systems with event-sampled state. For uncertain nonlinear systems with triangular structure a singularity-free adaptive fuzzy backstepping control is considered in [16]. Besides, this method is also used for inverted pendulum in some cases. For example, in [17], a model-free backstepping control technique is studied for a rotary inverted pendulum. [18] Presents a novel backstepping linear quadratic Gaussian controller based on the Lyapunov function for an inverted pendulum.

One of the most important issues in control theory is the actuator limitation which is significant from an implementation side. Besides, most of the control strategies are applied to this important issue [19]. In this study, a novel method is derived, in which, the backstepping approached is employed as a toll to solve the constrained optimization problem. On the other hand, the control constraint can be satisfied beside of performance goals, tracking the desired position. In addition, the minimum control effort is considered as a problem in this paper.

By changing the performance index to a constrained quadratic programming (QP) problem, the dynamic model of the projection recursive neural network (PRNN) is satisfied using variational inequality and projection theories. PRNNs are inspired by the Hopfield network where the performance index is minimized by converging the network to its equilibrium point. This method that a cost function is optimized by a neural network caused the appearance of a new dynamic structure named PRNNs, which are gained based on the variational inequality (VI) problem [20]. The VI problem is used to derive the dynamics of a PRNN and is extended to projection theory. These networks at first were used to solve linear programming problems [21]. Later, they were used for solving various QP [22]. nonlinear programming, and min-max optimization [23] and [24]. By simple structure and easily implemented, PTNNs show an effective analytical basis, and their fast convergence has led to development in real-time applications. PRNNs have been used widely in many areas like robotic [25] and predictive control [26].

So contributions of this paper are: By resorting to above mentioned issues, it is clear that many problems are remained unsolved and approaches like feedback controls such as conventional backstepping methods do not consider the actuator limitations and minimum control cost problem. additionally, most of the papers have been studied based on the linear or identified model of single inverted pendulum. Therefore, to solve this problem, an optimal backstepping control approach based on projection recurrent neural network is studied in this work for inverted pendulum. Here, a multi-loop strategy is proposed, where a constrained optimization problem is considered for each loop. By converting these problems into QP and considering control limitations, dynamic and output equations of the RNN are derived. This optimizer with simple structure updates the control signal simultaneously. Moreover, the convergence analysis of the neural optimizer and closed-loop stability have been performed.

## System description

The mathematical dynamic model of the single inverted pendulum according to the Fig. 1 is described as [27]:

$$\begin{cases} \dot{x}_1 = x_2 \\ \dot{x}_2 = \frac{g\sin(x_1) - \frac{mlx_2^2 \cos(x_1)\sin(x_1)}{(m_c + m)}}{l\left[\frac{4}{3} - \frac{m\cos^2(x_1)}{(m_c + m)}\right]} + d \\ \qquad + \frac{\frac{\cos(x_1)}{(m_c + m)}}{l\left[\frac{4}{3} - \frac{m\cos^2(x_1)}{(m_c + m)}\right]} * u \\ y = x_1 \end{cases} \quad (1)$$

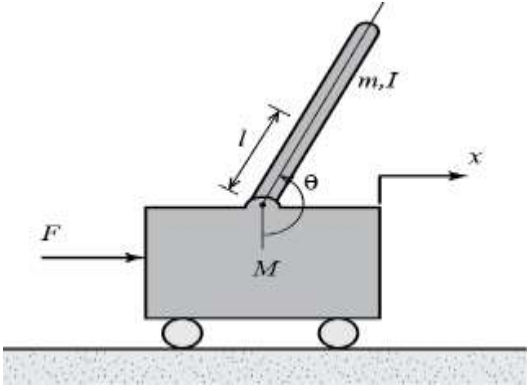

Fig 1. Schematic view of a magnetic levitation system

the pendulum parameters are shown in Table 1:

| | | |
|---|---|---|
| $g$ | gravitation acceleration | $9.8 m/s^2$ |
| $m_c$ | mass of the cart | $1 kg$ |
| $m$ | mass of the pendulum | $0.1 kg$ |
| $l$ | length to pendulum center of mass | $0.5 m$ |

Table 1 Single inverted pendulum

## Constrained quadratic problem adaptive backstepping control

The structure of backstepping control as an optimal constrained control is proposed in this section. As it was presented in the previous section, the dynamic model of the system is:

$$\begin{cases} \dot{x}_1 = x_2 \\ \dot{x}_2 = \dfrac{g\sin(x_1) - \dfrac{mlx_2^2 \cos(x_1)\sin(x_1)}{(m_c+m)}}{l\left[\dfrac{4}{3} - \dfrac{m\cos^2(x_1)}{(m_c+m)}\right]} + d \\ \quad + \dfrac{\dfrac{\cos(x_1)}{(m_c+m)}}{l\left[\dfrac{4}{3} - \dfrac{m\cos^2(x_1)}{(m_c+m)}\right]} * u \\ y = x_1 \end{cases} \quad (2)$$

for convince we consider two part of the dynamic as,

$$A = \dfrac{g\sin(x_1) - \dfrac{mlx_2^2 \cos(x_1)\sin(x_1)}{(m_c+m)}}{l\left[\dfrac{4}{3} - \dfrac{m\cos^2(x_1)}{(m_c+m)}\right]}$$

$$B = \dfrac{\dfrac{\cos(x_1)}{(m_c+m)}}{l\left[\dfrac{4}{3} - \dfrac{m\cos^2(x_1)}{(m_c+m)}\right]}$$

so the dynamic model of the system will be:

$$\begin{cases} \dot{x}_1 = x_2 \\ \dot{x}_2 = A + B*u + d \\ y = x_1 \end{cases} \quad (3)$$

so, based on the backstepping method, the virtual variable is considered as,

$$S_1 = x_1 - x_{1d} \quad (4)$$

$x_{1d}$ is desired trajectory of the inverted pendulum. we have:

$$\dot{S}_1 = x_2 - \dot{x}_{1d} \quad (5)$$

second variable is considered for stabilizing (5),

$$S_2 = x_2 - \dot{x}_{1d} - \gamma_1 \quad (6)$$

$\gamma_1$ is the first virtual control law and stabilizes the first variable. Therefore, we have:

$$\dot{S}_1 = S_2 + \gamma_1 \quad (7)$$

here, we need to have the Lyapunov function to prove the stability of the first virtual variable:

$$V_1 = \tfrac{1}{2}S_1^2 \quad (8)$$

if we take the time derivative we have:

$$\dot{V}_1 = S_1 S_2 + S_1 \gamma_1 \quad (10)$$

substituting $\gamma_1 = -c_1 S_1$ into (10). So we have:

$$\dot{V}_1 = S_1 S_2 - c_1 S_1^2 \quad (11)$$

if $S_2$ converges into zero, $V_1$ will be negative. Therefore, the dynamics of the second virtual variable is calculated as

$$\dot{S}_2 = A + B*u - \ddot{x}_{1d} - \dot{\gamma}_1 \quad (12)$$

by considering $\dot{\gamma}_1 = -c_1 S_2 + c_1^2 S_1$, (12) is:

$$\dot{S}_2 = A + B*u - \ddot{x}_{1d} + c_1 S_2 - c_1^2 S_1 \quad (13)$$

The Lyapunov function to prove the stability of the second virtual variable is considered as,

$$V_2 = \tfrac{1}{2}S_1^2 + \tfrac{1}{2}S_2^2 \quad (14)$$

if we take the time derivative we have:

$$\dot{V}_2 = S_1 S_2 - c_1 S_1^2 + S_2(A + B*u - \ddot{x}_{1d} + c_1 S_2 - c_1^2 S_1) \quad (16)$$

the equation (16) is equivalent as,

$$\dot{V}_2 = -c_1 S_1^2 + S_2(A + B*u - \ddot{x}_{1d} + c_1 S_2 + (1 - c_1^2)S_1) \quad (17)$$

by considering $A + B*u - \ddot{x}_{1d} + c_1 S_2 + (1 - c_1^2)S_1 = -c_2 S_2$, the required condition for stabilizing the second virtual variable is obtained and we have:

$$\dot{V}_2 = -c_1 S_1^2 - c_2 S_2^2 \leq 0 \quad (18)$$

To design a constrained optimal backstepping controller, the performance index is proposed as:

$$J = \frac{1}{2}T(A + B*u - \ddot{x}_{1d} + (c_1 + c_2)S_2 + (1 - c_1^2)S_1)^2 + \frac{1}{2}Ru^2 \quad (19)$$

here, $T$ and $R$ are weights of minimum tracking error and minimum control effort, respectively. The equation (19) will be:

$$J = \frac{1}{2}TB^2 u^2 + \frac{1}{2}T(A - \ddot{x}_{1d} + (c_1 + c_2)S_2 + (1 - c_1^2)S_1)^2 + TB(A - \ddot{x}_{1d} + (c_1 + c_2)S_2 + (1 - c_1^2)S_1)u + \frac{1}{2}Ru^2 \quad (20)$$

and equation (20) can be written:

$$J = \frac{1}{2}(TB^2 + R)u^2 + TB(A - \ddot{x}_{1d} + (c_1 + c_2)S_2 + (1 - c_1^2)S_1)u + \frac{1}{2}T(A - \ddot{x}_{1d} + (c_1 + c_2)S_2 + (1 - c_1^2)S_1)^2 \quad (21)$$

As it is evident, the third part of the performance index is independent of the control signal. So, by convergence of the mentioned term to zero, the convergence of the internal control law to its optimal point will be obtained. The optimal constrained optimization problem follows:

$$\min_{Subject\ to\ u_{min} \leq u \leq u_{max}} J = \frac{1}{2}(TB^2 + R)u^2 + +TB(A - \ddot{x}_{1d} + (c_1 + c_2)S_2 + (1 - c_1^2)S_1)u \quad (22)$$

If we consider two terms $P(x)$ and $Q(x)$ as:

$$P(x) = TB(A - \ddot{x}_{1d} + (c_1 + c_2)S_2 + (1 - c_1^2)S_1)$$

$$Q(x) = (TB^2 + R)$$

We would have:

$$\min_{Subject\ to\ u_{min} \leq u \leq u_{max}} J = \frac{1}{2}Q(x)u^2 + P(x)u \quad (23)$$

the equation (23) is the QP problem, and since $Q(x)$ is positive, (23) represents a convex constrained space. By the projection recurrent neural network, equation (23) is solved. This is proposed in the next section.

**Remark 1:** There were some parameters in the previous sections like $c_1, c_2, R, and\ T$, these parameters are represented as weights in the control system and have a key roles to reach the perfect results. Besides, it is evident that $T$ is larger than $R$ since $R$ is considered for avoiding numerical instability and $T$ is for improving the convergence rate of the optimizer.

## Using PRNN as a neural optimizer

The neural networks are usually used to learn the dynamic and behavior of the system. But here, by proposing a novel algorithm, PRNN is used as an approach to solve the constrained optimization problem. So, by combination of a first order dynamic equation and an algebraic equation representing a real-time optimizer, the PRNN is defined to solve the constrained QP problem (23). $f(u): R \to R$ is a continuous function and the gradient of the cost function and $\Psi = \{u \in R | u_{min} \leq u \leq u_{max}\}$ the constrained space, respectively, $u^*$ is the optimal solution of QP problem provided a solution for the following inequality is existed [28].

$$f(u^*) = (u - u^*) \geq 0 \quad \forall u \in \Psi \quad (24)$$

(24) is variational inequality (VI). According to (23), the gradient of the cost function is:

$$f(x, u) = Q(x)u + P(x) \quad (25)$$

due to fixed point theory, the optimal solution of the VI in (24) corresponds to the following projection problem [28].

$$PR_\Psi(u - f(x, u)) = u \quad (26)$$

that $PR_\Psi(u) = arg\ \min_{\xi \in \Psi}\|u - \xi\|$

$\xi$ is a point in the constrained space $\Psi$ and has the least distance from $u$ and $PR_\Psi(u)$ performs the

mapping of $u$ to a feasible space. (23) is converted to the following problem [29].

$$\max_{u(t)} J_{dual} = -\frac{1}{2}Q(x)u^2 + P(x)u + \mu u_{min} - \eta u_{max} \tag{27}$$

$$s.t. \quad Q(x)u(t) + P(x) - \mu + \eta = 0$$

where $\mu, \eta$ are the decision variables. By considering $\mu - \eta = \phi$ we will have:

$$Q(x)u(t) + P(x) - \phi = 0 \tag{28}$$

by resorting to the $\Psi$ we have:

$$\begin{cases} u_{min} & if \ u < u_{min} \\ u_{max} & if \ u > u_{max} \\ u & if \ u_{min} \leq u \leq u_{max} \end{cases} \tag{29}$$

so (29) corresponds to:

$$u(t) = PR_\Psi(u(t) - \phi(t)) \tag{30}$$

satisfying (30) is satisfying the projection equation (26). So, causes solving the optimization problem (23). Therefore, the dynamic model of PRNN is:

$$\vartheta \frac{d\phi}{dt} = PR_\Psi(u(t) - \phi(t)) - u(t) \tag{31}$$

where $\phi \in R$ and $\vartheta > 0$ are the state and convergence rate of PRNN, respectively.

**Remark 2**: The stability analysis of the PRNN assures that the dynamic in (31) asymptotically converges to its equilibrium point. This problem is similar to the concept of training of the Hopfield network.

**Remark 3**: It is worthwhile to mention that the complexity of the constraints in the optimization problem specifies the complexity of the PRNN. Hence, the complexity of the NN is based on the number of constraints.

**Remark 4**: Here the neural optimizer is considered to solve a constrained optimization problem. But it has the algebraic equation for $u(t)$. Under the condition of stability of the NN dynamics, this equation is used for closed-loop stability analysis. This feature cannot be found in other algorithms considered to solve constrained optimization problems.

**Remark 5**: According to the block diagram, the activation function of the NN is linear. Besides, the NN is a simple approach and these features make the NN perfect option for constrained optimization problems.

In coordination with the equilibrium point of the NN, the $u(t)$ converges to the optimal point. Due to (28), the $u(t)$ is gained as,

$$u(t) = Q(x)^{-1}(\phi(t) - P(x)) \tag{32}$$

We had the following dynamic model:

$$\dot{x}_2 = \frac{gsin(x_1) - \frac{mlx_2^2 \cos(x_1)\sin(x_1)}{(m_c + m)}}{l\left[\frac{4}{3} - \frac{mcos^2(x_1)}{(m_c + m)}\right]}$$
$$+ \frac{\frac{cos(x_1)}{(m_c + m)}}{l\left[\frac{4}{3} - \frac{mcos^2(x_1)}{(m_c + m)}\right]} * u \tag{33}$$

the (33) can be written as:

$$\dot{x}_2$$
$$= \frac{\frac{g(m_c + m)sin(x_1) - mlx_2^2 \cos(x_1)\sin(x_1) + cos(x_1)u}{(m_c + m)}}{\frac{\frac{4}{3}l(m_c + m) - mlcos^2(x_1)}{(m_c + m)}} \tag{34}$$

by applying some mathematical operations, it is equivalence to:

$$\frac{4}{3}l(m_c + m)\dot{x}_2 - ml\dot{x}_2 cos^2(x_1)$$
$$= g(m_c + m)sin(x_1)$$
$$- mlx_2^2 \cos(x_1)\sin(x_1) + cos(x_1) \rightarrow$$

$$\frac{4}{3}l(m_c + m)\dot{x}_2 = ml\dot{x}_2 cos^2(x_1) + g(m_c + m)sin(x_1)$$
$$- mlx_2^2 \cos(x_1)\sin(x_1) + cos(x_1) \rightarrow$$

$$\dot{x}_2 = \frac{3}{4}\frac{m}{(m_c + m)}\dot{x}_2 cos^2(x_1) + \frac{3}{4}\frac{g}{l}sin(x_1)$$
$$- \frac{3}{4}\frac{m}{(m_c + m)}x_2^2 \cos(x_1)\sin(x_1)$$
$$+ \frac{3}{4}\frac{cos(x_1)u}{l(m_c + m)}$$

so, the final term can be written as:

$$\dot{x}_2 = \frac{3}{4}\frac{m}{(m_c + m)}(\dot{x}_2 cos^2(x_1) - x_2^2 \cos(x_1)\sin(x_1))$$
$$+ \frac{3}{4}\frac{g}{l}sin(x_1) + \frac{3}{4}\frac{cos(x_1)u}{l(m_c + m)} \tag{35}$$

to improve the adaptability of the algorithm, parameter estimation using recursive least square is assumed as follow,

$$y = \Pi \theta^T \tag{36}$$

where

$$y = \dot{x}_2$$

$$\Pi = \left[\frac{3}{4}\dot{x}_2 cos^2(x_1) - \frac{3}{4}x_2^2 \cos(x_1)\sin(x_1)\right.$$

$$\left.\frac{3}{4}g\sin(x_1) \quad \frac{3}{4}\cos(x_1)u\right]$$

$$\theta = \left[\frac{m}{(m_c+m)} \quad \frac{1}{l} \quad \frac{1}{l(m_c+m)}\right]$$

the adaptation error is:

$$e_t = y - \Pi \hat{\theta}^T \tag{37}$$

by using the following recursive relations, unknown parameters will be:

$$\hat{\theta}(k+1) = \hat{\theta}(k) + G(k)e_t(k) \tag{38}$$

which we have:

$$G(k) = M(k-1)\Pi(k)[1 + \Pi^T(k)M(k-1)\Pi(k)]^{-1} \tag{39}$$

$$M(k) = \left(1 - G(k)\Pi^T(k)\right)M(k-1) \tag{40}$$

here, $G(k) \in R$ and $\Pi(k) \in R$ are error correction gain and the parameter estimation error covariance matrix, respectively.

the PRNN-ABC is updated by the estimated parameters of the inverted pendulum system. It means that,

$$u(t) = \hat{Q}(x)^{-1}\left(\phi(t) - \hat{P}(x)\right) \tag{41}$$

in which,

$$\begin{cases} P(x) = TB(A - \ddot{x}_{1d} + (c_1+c_2)S_2 + (1-c_1^2)S_1) \\ Q(x) = (TB^2 + R) \end{cases} \tag{42}$$

in other words,

$$\begin{cases} \hat{P}(x) = T\left(\dfrac{\dfrac{\cos(x_1)}{(\widehat{m_c}+\widehat{m})}}{\hat{l}\left[\dfrac{4}{3} - \dfrac{\widehat{m}\cos^2(x_1)}{(\widehat{m_c}+\widehat{m})}\right]}\right) \times \\ \qquad \left(\left(\dfrac{g\sin(x_1) - \dfrac{\widehat{m}\hat{l}\dot{x}_2^2 \cos(x_1)\sin(x_1)}{(\widehat{m_c}+\widehat{m})}}{\hat{l}\left[\dfrac{4}{3} - \dfrac{\widehat{m}\cos^2(x_1)}{(\widehat{m_c}+\widehat{m})}\right]}\right)\right. \\ \qquad \left. - \ddot{x}_{1d} + (c_1+c_2)S_2 + (1-c_1^2)S_1\right) \\ \hat{Q}(x) = \left(T\left(\dfrac{\dfrac{\cos(x_1)}{(\widehat{m_c}+\widehat{m})}}{\hat{l}\left[\dfrac{4}{3} - \dfrac{\widehat{m}\cos^2(x_1)}{(\widehat{m_c}+\widehat{m})}\right]}\right)^2 + R\right) \end{cases}$$

based on previous sections, now we are going to prove some theorems and Lemmas regarding the stability and convergence of the proposed approach.

**Theorem 1**: The dynamics model of PRNN converges to its equilibrium points due to the following proof.

**Proof**: We must consider a suitable Lyapunov function and optimal value $\phi^*$.

$$v(t) = \frac{\delta}{2}(\phi-\phi^*)^2 + \frac{\delta}{2}\hat{Q}(x)^{-1}(\phi-\phi^*)^2 \tag{43}$$

by time derivative we have:

$$\dot{v}(t) = \delta(\phi-\phi^*)\dot{\phi} + \delta\hat{Q}(x)^{-1}(\phi-\phi^*)\dot{\phi} \tag{44}$$

by applying the NN model we have:

$$\dot{v}(t) = (\phi-\phi^*)\left(PR_\Psi(u(t)-\phi(t)) - u(t)\right) \\ + \hat{Q}(x)^{-1}(\phi \\ -\phi^*)\left(PR_\Psi(u(t)-\phi(t)) \\ -u(t)\right) \tag{45}$$

by considering $u(t) - \phi(t) = \varpi$ and $u(t) \coloneqq \sigma$ we have:

$$\dot{v}(t) = (\phi-\phi^*)(PR_\Psi(\varpi)-\sigma) \\ + \hat{Q}(x)^{-1}(\phi-\phi^*)(PR_\Psi(\varpi) \\ -\sigma) \tag{46}$$

as it is clear, $\sigma$ is the control signal and is similar to variable e.

**Lemma 1**: For $PR_\Psi$, in space $\Psi$, there exist $\varpi$ and $\sigma$, such that the following is holds [28]:

$$(PR_\Psi(\varpi) - \sigma)(\varpi - PR_\Psi(\varpi)) \geq 0 \tag{47}$$

**Lemma 2**: For $u^*$ and $\phi^*$ the following inequality satisfies:

$$(\varrho - u^*)\phi^* \geq 0 \quad \forall \varrho \in \Psi \quad (48)$$

**Proof**: We had $Q(x)u(t) + P(x)$. So, $\phi^* = Q(x)u^*(t) + P(x)$. Additionally, (48) is a VI problem. Therefore, By Lemmas 1 and 2, (47) can be written:

$$(PR_\Psi(\varpi) - u^*)(\phi^* + \varpi - PR_\Psi(\varpi)) \geq 0 \quad (49)$$

we can write:

$$(PR_\Psi(\varpi) - u^* + u - u)(\phi - \phi^* + PR_\Psi(\varpi) - u) \leq 0 \quad (50)$$

due to (41) it can be written as:

$$(u - u^*) = Q(x)^{-1}(\phi - \phi^*) \quad (51)$$

and (50) can be stated as:

$$(PR_\Psi(\varpi) - u)(\phi - \phi^*) + (PR_\Psi(\varpi) - u)(u - u^*)$$
$$\leq |PR_\Psi(\varpi) - u|^2$$
$$- Q(x)^{-1}|\phi - \phi^*|^2 \leq 0 \quad (52)$$

In **Theorem 1**, the main consideration is: the state of the PRNN are differentiable, that due to (31) it is differentiable. The second assumption is that, the instantaneous data of the state variables, the states of the PRNN, and the reference input have to be available. Besides, (31) must be continuous and gradient must be achievable.

## Stability analysis of the system

Here, we prove that by small values for the weights of the control effort and positive values for the virtual control law, the closed-loop system is stable and it depends on the conditions of the PRNN.

**Theorem 2**, If:

(a) The dynamic model of the PRNN in (31) converges to zero.

(b) The weight $R$ is selected positive and close to zero,

then, the closed-loop system is asymptotically stable.

**Proof**: If $u_{min} \leq u \leq u_{max}$, the projection operator performs a mapping to its own argument:

$$PR_\Psi(u(t) - \phi(t)) = u(t) - \phi(t) \quad (53)$$

If $u_{min} \leq u \leq u_{max}$

the dynamic of the NN can be written:

$$\frac{d\phi}{dt} = -\vartheta\phi(t) \quad (54)$$

So we have:

$$\phi(t) = \phi_0 e^{-\vartheta t} \quad (55)$$

(55) proves that the state variable of the network converges to the origin from any initial condition if a positive learning rate $\vartheta$ is chosen and the size of the $\vartheta$ has direct relation with speed of convergence. The larger $\vartheta$, the faster speed of convergence of the network state variable to the origin.

It is shown that the inverted pendulum is asymptotically stable. Consider the Lyapunov function:

$$v = \frac{1}{2}z_1^2 + \frac{1}{2}z_2^2 \quad (56)$$

by applying derivation and substituting term we have:

$$\dot{v} = -c_1 S_1^2 + S_2(A + B * u - \ddot{x}_{1d} + c_1 S_2 + (1 - c_1^2)S_1) \quad (57)$$

as we had $u(t) = Q(x)^{-1}(\phi(t) - P(x))$:

$$u(t) = Q(x)^{-1}\phi(t) + Q(x)^{-1}(-TB(A - \ddot{x}_{1d} + (c_1 + c_2)S_2 + (1 - c_1^2)S_1)) \quad (58)$$

by using (57) and (58) we have:

$$\dot{v} = -c_1 S_1^2 + S_2(A - \ddot{x}_{1d} + c_1 S_2 + (1 - c_1^2)S_1)$$
$$+ S_2(BQ(x)^{-1}\phi(t))$$
$$+ S_2\left(Q(x)^{-1}(-TB^2(A - \ddot{x}_{1d} + (c_1 + c_2)S_2 + (1 - c_1^2)S_1))\right) \quad (59)$$

if we assume:

$$1 - TB^2 Q(x)^{-1} = 0 \quad (60)$$

we can write the equation (59):

$$\dot{v} = -c_1 S_1^2 - c_2 S_2^2 + S_2(BQ(x)^{-1}\phi(t)) \quad (61)$$

due to (55), $\lim_{t \to 0} BQ(x)^{-1}\phi(t) \to 0$

therefore, by choosing the positive values for $c_1$ and $c_2$, the equation (61) is negative.

## Conclusion

In this paper, a novel recurrent adaptive backstepping optimal control strategy for tracking a single inverted pendulum system is studied. Here, the design of backstepping control and a performance index according to constraints are considered, and to solve the constrained optimization problem, the dynamic model of the PRNN is assumed. Convergence and stability analysis is done due to the Lyapunov theorem. In this paper, main considerations are actuator limitation, minimizing control effort, simple and online optimization problem, adaptability of parametric variations. performance speed, no overshoot, and no need for changes in the control signal to improve its convergence speed are another advantages of the proposed algorithm.